\documentclass[12pt]{article}
\usepackage{lscape,slashed}
\usepackage{multirow}
\usepackage{color}
\usepackage{bm}
\usepackage{comment}
\usepackage{amssymb,amsmath,yhmath,mathrsfs,graphicx}
\usepackage{amstext}
\usepackage{amscd}
\usepackage{graphics}
\usepackage{epsfig}
\usepackage{shadow}
\usepackage[all]{xy}
\usepackage{hyperref}

\def\sss{\scriptscriptstyle}

\newcommand{\eqn}[1]{(\ref{#1})}

\def\sss{\scriptscriptstyle}

\def\be{\begin{equation}}
\def\ee{\end{equation}}
\def\bes{\begin{equation*}}
\def\ees{\end{equation*}}
\def\beq{\begin{equation}}
\def\eeq{\end{equation}}
\def\bea{\begin{eqnarray}}
\def\eea{\end{eqnarray}}
\def\beas{\begin{eqnarray*}}
\def\eeas{\end{eqnarray*}}

\def\nn{\nonumber}

\def\sideremark#1{\ifvmode\leavevmode\fi\vadjust{\vbox to0pt{\vss
 \hbox to 0pt{\hskip\hsize\hskip1em
 \vbox{\hsize2cm\tiny\raggedright\pretolerance10000
  \noindent #1\hfill}\hss}\vbox to8pt{\vfil}\vss}}}

\begin{document}
\thispagestyle{empty}

\setcounter{footnote}{0}

\vspace{-50mm}
\begin{flushright}\tt CALT-TH 2014-150\end{flushright}

\begin{center}

{\Large
 {\bf Covariant constraints for generic massive gravity and~analysis of its characteristics}\\[4mm]

 {\sc \small
    S.~Deser$^{\mathfrak C}$,  M.~Sandora$^{\mathfrak{G}, \mathfrak{D}}$, A.~Waldron$^{\mathfrak M}$ and G. Zahariade$^{\mathfrak D}$}\\[3mm]

{\em\small
                      ~\ ${}^\mathfrak{C}$
                       Walter Burke Institute for Theoretical Physics, California Institute of Technology, Pasadena, CA 91125; 
     Physics Department, Brandeis University, Waltham, MA 02454.\\
{\tt deser@brandeis.edu}\\[2mm]
          
           ~${}^{\mathfrak G}\!$
           CP3-Origins, Center for Cosmology and Particle Physics Phenomenology,
University of Southern Denmark, Campusvej 55, 5230 Odense M, Denmark
           \\[2mm]

           ~${}^{\mathfrak D}\!$
            Department of Physics\
            University of California,
            Davis CA 95616, USA\\
            {\tt mesandora@ucdavis.edu, zahariad@ucdavis.edu}\\[2mm]
          
       ~${}^{\mathfrak M}\!$
            Department of Mathematics\
            University of California,
            Davis CA 95616, USA\\
            {\tt wally@math.ucdavis.edu}

            }
}

\bigskip

{\sc Abstract}\\[1mm]

\end{center}

\noindent
We perform a covariant constraint analysis of massive gravity valid for its entire parameter space,  demonstrating that
 the model generically propagates five degrees of freedom; this is also verified by a new and streamlined Hamiltonian description.  The  constraint's   covariant  expression permits computation of  the model's caustics. Although new features such as 
the dynamical Riemann  tensor appear in the characteristic matrix, 
  the model still exhibits  the pathologies uncovered in earlier work: superluminality and likely acausalities. 

\newpage



\section{Introduction}

Massive gravity (mGR) models defined in terms of a fiducial metric  have been intensely studied in recent years in the hope of providing an observationally viable, finite range, extension of Einstein's general relativity (GR)~\cite{dRGT}.
This spate of activity occurred despite the fact that no definitive analysis of fiducial massive gravity (fmGR\footnote{ These models have also been dubbed ``dRGT mGR'' or ``ghost-free mGR''. We prefer the more descriptive fmGR title, because these models are not ghost-free (see below) but rather only avoid the obvious sixth, zero helicity, ghostlike excitation. Moreover their inconsistencies can be traced back to the external-fiducial background.}) propagation and causal properties valid for its  full parameter range had 
been undertaken; this is our aim: Our  findings bolster earlier ones of both acausality and superluminality.
The key technical advance enabling these computations is the first covariant degree of freedom (DoF)  analysis valid for the model's full parameter range. We will also present an improved Hamiltonian analysis as a check on these findings.

It was realized  long ago that interacting higher spin $s\geq 1$ fields can suffer from a variety  of inconsistencies. The first issue is that the field theoretic propagating DoF of the interacting theory may not match those of its free limit. As shown in~\cite{BD}, generic massive gravity theories fail at  this first hurdle. However, even models 
passing this first consistency barrier---in particular fmGR---may still propagate unphysical modes. This phenomenon was first observed in the context of the canonical quantum commutators of charged spin 3/2 fields; they  were found to be pathological in EM backgrounds~\cite{KS}. That  pathology was later traced  back to  the underlying kinetic structure of the theory. The latter was  studied  by searching for superluminal shock wave solutions to the underlying PDEs~\cite{VZ} and extended to spin~2 in~\cite{KoS}. Shock waves propagate on characteristic surfaces, off of which  the evolution of all physical variables is no longer determined. This explains why zero and negative norm states appear in canonical commutators. In background-independent GR, the characteristic surfaces encode the causal structure of the theory and are not fatal {\it per se}. However, if one takes this viewpoint (thus abandoning fmGR consistency as a spin~2 field theory in its fiducial background), there remains the further requirement that solutions with local\footnote{The less pernicious global  CTCs of G\"odel type are in principle still permitted.} closed timelike curves (CTCs) be absent. These are notoriously difficult to avoid in models with field-dependent characteristic matrices~\cite{Nester}. 

The first fmGR model was given by Zumino  in 1970~\cite{Zumino}, by setting one of the two dynamical metrics of the, then new, 
 bimetric ``$f$-$g$'' theories of Isham, Salam and Strathdee~\cite{Salam}  to a fixed (fiducial) background and requiring the free limit to be the massive, $s=2$ Fierz--Pauli (FP) theory. 
However, it was soon realized~\cite{BD} that mGR models generically included an additional, sixth, ghost-like, zero helicity, field theoretic DoF. Furthermore, even (linear) FP theory was found to predict incorrect results for bending of light in its vanishing mass limit~\cite{vDVZ}. Subsequently it was argued that this difficulty could be an artefact of the linearized limit--setting the interaction strength to zero before the massless limit could cause the  faulty light-bending predictions~\cite{Va}. Alas, in the absence of a consistent interacting massive model, this suggestion was very difficult to verify (although it was shown that a similar mechanism for the  FP model in cosmological backgrounds, interchanging limits of vanishing mass and cosmological constant did cure the light-bending disease~\cite{Lambda}). This set the stage for effective field theorists to apply the decoupling limit (large Planck mass~$M_{\rm P}$, small graviton mass~$m$ and constant~$m^2 M_{\rm P}$) technique to study fmGR's dangerous zero helicity sector. Remarkably, they recovered  Zumino's original fmGR model plus two further extensions   as candidate ghost-free theories~\cite{dRGT}. 

At this point, a frenzy of mGR activity ensued (see the reviews~\cite{HinterbichlerReview}); but some darker clouds had gathered on the horizon:
An intricate, $(3+1)$,  ADM constraint analysis verified that the fmGR models propagated five field theory DoF but cast little light on its kinetic structure~\cite{HR}, except that it was rather complicated--to be precise, various implicit field redefinitions were needed, yielding a potentially pathological symplectic current. Indeed, already in the decoupling limit superluminalities had been detected~\cite{Gr}. This indicated that the difficulties faced by other (finite tower) higher spin models would likely befall fmGR. Indeed,  a second order shock analysis discovered fmGR superluminalities, at least for a one-dimensional subspace of its allowed parameter values~\cite{DW}; this result was extended to a two-dimensional subspace in~\cite{DSW}. These were later shown to be  consequences of superluminal behavior detected via a first order computation of the model's characteristic matrix~\cite{DeserOng} (see also~\cite{Izumi,Problems}). Worse still, this first order computation showed how to use superluminality to embed closed timelike loops and thus violate microcausality. Hand in hand with those results, it was also discovered that fmGR possessed no consistent partially massless limit~\cite{DSW,deRhamPM}. [Since partially massless theories were originally discovered by demanding lightlike propagation~\cite{PM}, and underlie the cosmological solution to the light-bending problem~\cite{Lambda}, this constitutes strong evidence against fmGR consistency.] 

In this article we extend earlier constraint and propagation analyses to the full fmGR parameter range. The original covariant analysis of constraints in vierbein form~\cite{DMZ} pioneered this approach for two of the three allowed mass terms, albeit failing when applied to the remaining direction in the fmGR parameter space. This was because---seemingly non-removable---terms appeared in the putative scalar constraint that involved the full dynamical Riemann tensor. Thus, given that previous (3+1) constraint analyses for this case were rather implicit~\cite{HR}, absence of the field theoretic ghost in this corner of fmGR parameter space could not be fully confirmed. 
[Other groups have investigated the full parameter space, but only for specialized field configurations and agree with our result~\cite{Military}.]
Our aim is therefore to build upon the methods of~\cite{DMZ} to fill in this gap. This also allows us to compute the characteristic matrix for the full fmGR parameter space; indeed the formerly troublesome Riemann terms imply a new dependence of the characteristic matrix on dynamical  curvatures. 


The characteristic matrix is a powerful tool for examining consistency of models. Ultimately, fmGR proponents would need to show non-vanishing of its determinant to save the model from pathology. This criterion could be used both to discover a preferred parameter choice and to determine a preferred fiducial background. [The freedom to choose by hand the fiducial metric in order to fit data implies a massive loss of predictability.]
However, in~\cite{DeserOng} fatal acausalities for very general field configurations and independent of choice of fiducial metric were uncovered, so the range of physical viability of fmGR theories is likely to be highly limited at best.  In this article, we content ourselves with exhibiting the characteristic method 
at work for some simple examples around flat fiducial backgrounds.

Our article is structured as follows. Our covariant constraint analysis is given in Section~\ref{cca} and the characteristic matrix is computed in Section~\ref{cm}. We analyze the characteristic matrix for propagation pathologies in Section~\ref{pa}. Our conclusions, where we discuss  fmGR's last vestige of applicability as an effective field theory as well as related models such as the bimetric theory where the fiducial background is promoted to a dynamical field, are given in Section~\ref{c}. 
In Appendix~\ref{A} we present the linear limit of our first order, covariant constraint analysis, while Appendix~\ref{B} gives a rapid  sketch of the model's frame-like Hamiltonian, 
description from which the DoF count can also be checked.

\section{Covariant constraint analysis}\label{cca}

At their genesis, bimetric~\cite{Salam} and massive gravity~\cite{Zumino} were originally formulated in terms of vierbeine~$e^m$ and~$f^m$.
Both these fields are dynamical for the bimetric theory, while for~mGR,~$f^m$ is taken to be a fiducial background ({\it e.g.}, $\bar g_{\mu\nu}:=f_\mu{}^m\eta_{mn} f_\nu{}^n$). In these terms, the statement of 
the model and its constraint analysis are rather simple.

Throughout this Section, unless explicitly noted, we will use  a differential form notation where wedge products are assumed whenever obvious. The action is a sum of Einstein--Hilbert and mass terms:
\bes
{ S}_{\rm fmGR}:=S_{\rm GR}+{S}_{\rm fm}\ ,
\ees
where
\beas
S_{\rm GR}&:=&\,\, -\frac{1}{4}\, \int \epsilon_{mnrs}\,  e^m{{}} e^n \, \left[d\omega^{rs} + \omega^r{}_t{{}}\omega^{ts}\right]\, ,\\[3mm]
{S}_{\rm fm}&:=&\!\!\!\!\phantom{-}m^2 \int \epsilon_{mnrs}\, e{}^m{{}}\left[\frac{\beta_{0}}{4}e^n{{}} e^r{{}} e^s +\frac{\beta_{1}}{3}e^n{{}} e^r{{}} f^s+\frac{\beta_{2}}{2}e^n{{}} f^r{{}} f^s+\beta_{3}f^n{{}} f^r{{}} f^s\right]\, .
\eeas
Note that $\beta_0$ parameterizes a standard cosmological term (which will be required to obtain the FP linearized limit, even around flat, Minkowski, backgrounds), while a $\beta_4$ term made from four fiducial vierbeine only contributes an irrelevant additive constant,
so has been omitted. It is known that the model can be linearized around fiducial Einstein backgrounds with cosmological constant $\bar \Lambda$ only if  the model's parameters obey 
\be\label{backsol}
\frac{\bar \Lambda}{3!} = m^2 \left(\beta_0+\beta_1+\beta_2+\beta_3\right)\, .
\ee
Its linearized limit (see Appendix~\ref{A}) is then the FP theory with mass
\be\label{PauliFierz}m_{\rm FP}^2:=m^2(\beta_1+2\beta_2+3\beta_3)\, .\ee
The model's dynamical fields are the vierbein and spin-connection one-forms~$(e^m, \omega^{mn})$ whose variations
 give equations  of motion:
\bea
 {\cal T}^{m}&:=&\nabla e^{m}\approx 0\ ,\nn\\[2mm]
 {\cal G}_{m}&:=&G_{m}-m^{2}t_{m}\approx 0\ .\label{eoms}
\eea
Equations which hold on-shell are written using the weakly vanishing notation~$\approx\,$ and a calligraphic font will be used for weakly vanishing quantities.
The first equation implies vanishing torsion so that the spin-connection weakly equals the Levi-Civita one.  The second equation is the standard Einstein equation modified by the mass term.
 In the above, we have denoted the exterior covariant derivative with respect to~$\omega^{mn}$ by~$\nabla$ so that for any Lorentz vector-valued form~$\sigma^m$
$$
\nabla \sigma^m:=d \sigma^m + \omega^m{}_n \sigma^n\, . 
$$
Moreover we have defined the Einstein three-form 
 \be\label{Einstein3}
 G_{m}:=\frac{1}{2}\epsilon_{mnrs}e^{n}{{}} R^{rs}\, ,
 \ee
where the two-form~$R^{mn}:=d\omega^{mn}+\omega^{m}{}_{t}{{}}\omega^{tn}$ is the Riemann curvature associated to the connection.
The dual of the display~\eqn{Einstein3} is
the Einstein tensor. Finally the {\it mass stress-tensor} is encoded by the three-form~\cite{HBR}
\be\label{stews}
 t_{m}:=\epsilon_{mnrs}\, \left[\beta_{0}e^n{{}} e^r{{}} e^s +\beta_{1}e^n{{}} e^r{{}} f^s+\beta_{2}e^n{{}} f^r{{}} f^s+\beta_{3}f^n{{}} f^r{{}} f^s\right]\, .
 \ee

To analyze the model's constraints we need a notion of timelike evolution. For that, one assumes invertibility of the dynamical vierbein and in turn of the metric~$ds^2=e^m\otimes e_m$,
which is taken to have signature~$(-,+,+,+)$. [Our analysis easily extends to arbitrary dimensions $d\geq 3$, see footnote~\ref{ddim}.] 
Hence, for any choice of  timelike\footnote{Here we mean timelike with respect to the dynamical metric, although none of the constraints found in this Section depend essentially on the choice of foliation of the underlying spacetime manifold.} evolution parameter~$t$ we can decompose a~$p$-form~$\mbox{\resizebox{2.5mm}{2.85mm}{$\theta$}}$ (with~$p<4$) as
\bea\label{ring}
 \mbox{\resizebox{2.5mm}{2.85mm}{$\theta$}}:= \bm \theta+\mbox{\resizebox{2.5mm}{4.1mm}{$\ring\theta$}}\, ,
\eea
where~$\mbox{\resizebox{2.5mm}{4.1mm}{$\ring\theta$}}\wedge dt=0$. Thus~$\bm \theta$ is the purely spatial part of the form~\resizebox{2.5mm}{2.85mm}{$\theta$}. The beauty of this notation is that the purely spatial~$\bm {\mathcal P}\approx 0$ part of any on-shell relation~${\cal P}\approx 0$ polynomial in $(\nabla,e,\omega)$ is a constraint because it necessarily contains no time derivatives.
Our analysis proceeds along the same lines as in~\cite{DMZ,DW,DSW} albeit in a first-order formalism ``\`a la Palatini''. The forty first order equations of motion for forty fields ultimately describe (at least generically) ten propagating fields, so five physical DoF. To  establish this result in a simple covariant formalism, we therefore need to find thirty constraints, {\it i.e.}, weak relations not involving time derivatives of dynamical fields.
Sixteen of these are given directly  by the equations of motion themselves and are thus primary constraints. Evolving these gives ten secondary constraints whose evolution in turn yields the final four tertiary constraints.

 \subsection{Primary constraints}
 
 The spatial parts of the equations of motion~\eqn{eoms} give sixteen primary constraints
 \bea
{\bm {\mathcal T}}^{m}&\approx& 0\, ,\nn\\[3mm]
{\bm {\mathcal G}}_{m}&\approx& 0\ .\label{frogs}
\eea
In terms of dynamical fields, these read
\beas
&\bm\nabla \bm e^m:=\bm d \bm e^m + \bm \omega^m{}_n \bm e^n\approx 0\, ,\\[3mm]
&\frac{1}{2}\epsilon_{mnrs}\bm e^n \left(\bm d \bm \omega^{rs}+\bm \omega^n{}_t\bm \omega^{ts}\right)\approx
m^2 \epsilon_{mnrs}\left(\beta_0\bm e^n\bm e^r \bm e^s+
\beta_1\bm e^n\bm e^r \bm f^s+
\beta_2\bm e^n\bm f^r \bm f^s+
\beta_3\bm f^n\bm f^r \bm f^s\right)\, .
\eeas

\subsection{Secondary constraints}

In principle we could compute secondary constraints by brute force by taking a time derivative of the primary constraints~\eqn{frogs}. That computation is vastly simplified 
by considering their exterior covariant derivatives. The purely spatial part of this is of course not a new constraint, but the remainder, modulo the field equations
can possibly yield new, secondary, constraints.

\subsubsection{The symmetry constraint}

On-shell, $G_m\wedge e_{n}$ is equal to the volume form multiplied by the Einstein tensor and thus symmetric under interchange of $m$ and $n$.
Indeed,   
\bes
G_{[m}e_{n]}=\frac{1}{2}\epsilon_{mnrs}e^{r}{{}}\nabla{\cal T}^{s}\approx 0\ .
\ees
This leads to six secondary constraints
\bes
t_{[m}e_{n]}=\frac{1}{m^{2}}\left(\frac{1}{2}\epsilon_{mnrs}e^{r}{{}}\nabla{\cal T}^{s}-{\cal G}_{[m}{{}} e_{n]}\right)\approx 0\ .
\ees
Using the Schouten identity\footnote{\label{Schouten}This (tautological) identity states $v_m\epsilon_{nrs\ldots}=
v_n\epsilon_{mrs\ldots}+v_r\epsilon_{nms\ldots}+v_s\epsilon_{nrm\ldots}+\cdots$.}, this gives
\be
{M}^{mn}{{}}{\cal F}\approx 0\ ,
\label{symmetry}
\ee
in terms of the two-forms
\bes
{\cal F}:=e^{m}{{}} f_{m}\quad\text{and}\quad {M}^{mn}:=\beta_{1} e^{m}{{}} e^{n}+2\beta_{2} e^{[m}{{}} f^{n]}+3\beta_{3} f^{m}{{}} f^{n}\ .
\ees
In~\eqn{symmetry} the operator ${M}^{mn}$ maps two-forms to antisymmetric  Lorentz tensors (multiplied by the volume form)
and is therefore generically  invertible.
Thus, in the following we will assume
\be\label{symmetry1}{\cal F}\approx 0\ee (hence the calligraphics) even if this is technically not implied  for all regions of parameter space\footnote{For example, when~$\beta_{1}=\beta_{3}=0$ the operator above is not invertible. In the cases~$\beta_{2}=\beta_{3}=0$ and~$\beta_{1}=\beta_{2}=0$ as well as~$\beta_{2}=\beta_{1}\lambda$,~$\beta_{3}=\frac{\beta_{1}\lambda^{2}}{3}$ (so long as~$e^{m}+\lambda f^{m}$ is a basis of the cotangent space) equation~\eqn{symmetry} does imply ${\cal F}\approx0$. 
}~\cite{DeffayetBein}. We call this the \textit{symmetry constraint}. 

\subsubsection{The vector constraint}

The Einstein tensor's Bianchi identity  implies that  diffeomorphism invariant metric DoF must be coupled to divergence-free sources.
In the fmGR context, this yields a constraint. Here computing the covariant curl of the Einstein three-form, using $\nabla R^{mn}\equiv 0$, gives
\bes
\nabla G_{m}= \frac{1}{2}\epsilon_{mnrs} {\cal T}^{n}{{}} R^{rs}\approx0\ .
\ees
This leads to the four constraints
\be\label{prevector}
\nabla t_{m}=\frac{1}{m^{2}}\left(\frac{1}{2}\epsilon_{mnrs} {\cal T}^{n}{{}} R^{rs} - \nabla{\cal G}_{m}\right)\approx 0\ .
\ee
We can write these constraints explicitly because
\beas
\ \nabla t_{m}\!\!&\!\!=\!\!& \epsilon_{mnrs}{\cal T}^{n}{{}}\left(3\beta_{0}e^{r}{{}} e^{s}+2\beta_{1}e^{r}{{}} f^{s}+\beta_{2}f^{r}{{}} f^{s}\right)+\, \epsilon_{mnrs}{M}^{nr}{{}} K^{st}{{}} f_{t}\nn\\[2mm]
&\!\!=\!\!& \epsilon_{mnrs}\left[{\cal T}^{n}{{}}\left(3\beta_{0}e^{r}{{}} e^{s}+2\beta_{1}e^{r}{{}} f^{s}+\beta_{2}f^{r}{{}} f^{s}\right)+{\cal F}{{}}\left(\beta_{1}e^{n}+\beta_{2}f^{n}\right){{}} K^{rs}\right]-\frac{1}{2}\epsilon_{nrst}{M}^{nr}{{}} K^{st}{{}} f_{m}\ .
\eeas
Here we have defined the contorsion
$$
K^{m}{}_{n}:= \omega^{m}{}_{n}-\bar{\omega}^{m}{}_{n}\ ,
$$ 
where $\bar{\omega}^{m}{}_{n}$ is the fiducial Levi-Civita spin connection. The contorsion measures the difference between dynamical and fiducial spin-connections, thus
\bes
\nabla f^{m}=K^{m}{}_{n}{{}} f^{n}\ .
\ees
Hence, using invertibility of the fiducial vierbein, we finally\footnote{For the reason mentioned above, the equation ${\bm \nabla} {\bm {\mathcal T}}^{m}={\bm R}^{m}{}_{n}{{}} {\bm e}^{n}\approx 0$ yields no further secondary constraints since it is the spatial derivative of~${\bm {\mathcal T}}^{m}\approx 0$.} have the \textit{vector constraint}
\be
{\cal V}:=\epsilon_{mnrs}{M}^{mn}{{}} K^{rs}\approx 0\ .
\label{vector}
\ee

\subsection{Tertiary constraints}

We must now compute the time evolution of the ten secondary constraints, comprised of the six symmetry~\eqn{symmetry}, and four vector~\eqn{vector} constraints. This will lead, respectively, to three and one additional tertiary constraints.

\subsubsection{Evolving the symmetry constraint}

Since the symmetry constraint is the weak vanishing of the two-form ${\cal F}$, we can simply take the covariant curl of~\eqn{symmetry} to generate tertiary constraints: 
\bes
\nabla {\cal F} = {\cal T}^{m}{{}} f_{m} +K_{mn}{{}} e^{m}{{}} f^{n}\approx 0\, ,
\ees
which yield the three-form, {\it curled symmetry constraint}:
\be\label{Kef}
{\cal K}:=K_{mn}{{}} e^{m}{{}} f^{n}\approx 0\ .
\ee
This three-form might seem to constitute four new constraints, but exactly as above, its
purely spatial part  is just the spatial derivative of the symmetry constraint and hence not new. 
Therefore, in the notation of~\eqn{ring}, we have three new constraints 
\bes
\ring K^{mn} \bm e_m \bm f_n+\bm K^{mn} \ring e_m \bm f_n+\bm K^{mn} \bm e_m \ring f_n
\approx 0\ .
\ees

\subsubsection{Evolving the vector constraint }\label{evolution}

Since the vector constraint is the weak vanishing of the three-form ${\cal V}$, we take the curl of~\eqn{vector} to generate the final, scalar, tertiary constraint:
$$
\nabla {\cal V} = \epsilon_{mnrs}\left[{2\cal T}^{m}{{}}\left(\beta_{1} e^{n}+\beta_{2} f^{n}\right){{}} K^{rs}\!-\!\left(2\beta_{2}e^{m}+6\beta_{3}f^{m}\right){{}} K^{nr}{{}} K^{s}{}_{t}{{}} f^{t}+
{M}^{mn}
\nabla K^{rs}\right]\approx 0\ .
$$
At first glance, the above equation is not obviously a constraint because the last term, involving the curl of the contorsion, could  contain a time derivative of the dynamical spin connection. To see that this is not the case, we begin with an 
identity:
\be\label{dusttodust}
\nabla K^{mn}\equiv R^{mn}-\bar{R}^{mn}+K^{m}{}_{t}{{}} K^{tn}\ ,
\ee
where the two-form~$\bar{R}^{mn}$ is the fiducial Riemann curvature.
This shows that dangerous time derivatives can only arise via the dynamical Riemann curvature~$R^{mn}$. 
However, the equations of motion~\eqn{eoms} tell us that the Einstein tensor $G_{\mu\nu}$ 
 weakly equals terms involving no time derivatives (namely the mass stress tensor). Moreover, standard identities for the Riemann tensor all hold weakly, in particular
its divergence is related to the curl of the Einstein tensor by 
\be\label{divR}
\nabla^\mu R_{\mu\nu\rho\sigma}\approx 2\nabla_{[\rho}\Big( G_{\sigma]\nu}-\frac12 g_{\sigma]\nu}G_\mu{}^\mu\Big)\, .
\ee
Hence, on-shell,
the quantity $\partial_tR_{0\nu\rho\sigma}$ generically
has at most one time derivative on dynamical fields\footnote{In more detail, $\nabla^\mu R_{\mu\nu\rho\sigma}=\nabla^0 R_{0\nu\rho\sigma}+\cdots=g^{00}\dot R_{0\nu\rho\sigma}+\cdots$, where the ``$\cdots$'' terms involve at most one time derivative of the dynamical fields $(e^m,\omega^{mn})$.}, so  the only dangerous Riemann components $R_{0\nu\rho\sigma}$ have none and hence, in turn,  nor does the curvature $R^{mn}$. 
[This simple covariant argument is also readily verified by writing out the equations of motion~\eqn{eoms} in an explicit  $3+1$ split for any choice of time coordinate.]
 Therefore
 \be\label{scal}
{\cal S} := \epsilon_{mnrs}\Big[{M}^{mn}
\nabla K^{rs}-\left(2\beta_{2}e^{m}+6\beta_{3}f^{m}\right){{}} K^{nr}{{}} K^{s}{}_{t}{{}} f^{t}
\Big]\approx 0\ 
\ee
is a constraint equation. This is the long-sought {\it scalar constraint}.
In the special case $\beta_3=0$, the curl of the contorsion in the above display is traced
with the {\it dynamical} vierbein and thus can be converted to a  trace of the Riemann tensor,  {\it i.e.} the Einstein tensor, 
which can be handled directly by the equations of motion. In contrast, when $\beta_3\neq 0$, the dangerous Weyl part of Riemann is only traced with fiducial vierbeine 
so one must rely on equation~\eqn{divR} to prove that ${\cal S}$ is a constraint.
This explains why
 previous works~\cite{DMZ,DW,DSW,DeffayetBein} failed to find a covariant expression valid for the entire $(\beta_0,\beta_1,\beta_2,\beta_3)$ parameter space. 

 It will be useful to have a more explicit expression for the scalar constraint.
For that, we employ equations~(\ref{Einstein3},\ref{dusttodust}), and the Schouten identity (see footnote~\ref{Schouten}) to rewrite it as
\bea\label{scal1}
\mathcal{S}\!\!&=&\!\! \epsilon_{mnrs}\left(\beta_{1}e^{m}e^{t}-2\beta_{2}e^{(m}f^{t)}-3\beta_{3}f^{m}f^{t}\right)K^{nr}K^{s}{}_{t}\nn\\[2mm]
&&\quad\quad+\, 2\, \left(\, \beta_{1}e^{m}+2\beta_{2}f^{m}\right)G_{m}+3\epsilon_{mnrs}\beta_{3}f^{m}f^{n}R^{rs}\nn\\[2mm]
&&\quad\quad-\, 2\left(2\beta_{2}e^{m}+3\beta_{3}f^{m}\right)\bar{G}_{m}-\epsilon_{mnrs}\beta_{1}e^{m}e^{n}\bar{R}^{rs}\approx 0\ ,
\eea
where $\bar{G}_{m}:= \frac12 \epsilon_{mnrs}f^n \bar R^{rs}$ is the background Einstein tensor. 
Here one can exchange the dynamical Einstein tensor for the mass stress tensor $G_m\approx m^2 t_m$, whose explicit expression (algebraic in dynamical fields) is given in~\eqn{stews}. Moreover, remember that for the stubborn case $\beta_3\neq 0$, the term involving the Riemann tensor (weakly) does not depend   on time derivatives of dynamical variables. The above expression coincides with known results for the covariant scalar constraint for $\beta_3=0$~\cite{DW,DSW}. Also, specializing to the case $\beta_0=\beta_1=\beta_3=0$ and choosing the partially massless tuning of $\beta_2$ to the background cosmological constant~\cite{PM}, only the terms involving the square of the contorsion remain. These are precisely the obstruction to a partially massless limit of massive gravity~\cite{DSW,deRhamPM}.

At this point, so long as we can establish that the thirty constraints found so far are independent, the model describes no more than five physical DoF\footnote{\label{ddim} In $d\geq 3$ dimensions the model propagates $\frac12(d+1)(d-2)$ physical DoF, which can be seen as follows: There are $d^2+\frac12d^2(d-1)$ dynamical vielbeine and spin connections. These are subject to  $d+ \frac12 d(d-1)(d-2)$ primary,  $d+\frac12d(d-1)$ secondary, and  $1+\frac12(d-1)(d-2)$ tertiary constraints. This leaves $(d+1)(d-2)$ first order propagating fields which yields the quoted DoF count.}.
For the subspace of parameter space given by models which linearize to FP, this is essentially guaranteed  (see  Appendix~\ref{A}). The possibility that fewer DoF propagate, especially in special limits, such as the massless Einstein or a putative partially massless limit remains. The former of course, holds, but the latter possibility was ruled out in~\cite{DW,DSW}. It  could also be that, for parameter branches where the symmetry constraint 
is not guaranteed,  fewer DoF propagate.

\section{The Characteristic Matrix}\label{cm}

For a system of coupled, first order PDEs, we must ask whether, given initial data,  higher derivatives of fields are determined. This question is addressed
by  the system's characteristic matrix which can be computed by studying shocks~\cite{CourantHilbert,Problems}. In particular, the characteristic surface is defined by vanishing of the corresponding determinant.
Shocks propagate along this surface--where all higher derivatives can no longer be determined. In particular, spacelike characteristic surfaces signal superluminal shock propagation.
These foretell doom for the model viewed as a theory of a spin~2 field in a fiducial background. This method also allows us to study whether the model can access the escape route 
taken by (background independent) GR, whose characteristics determine its physical causal structure. 
Thus we now  study fmGR  shocks along surfaces with timelike normal vectors. These determine the model's  characteristic matrix from its equations of motion and constraints, as given in the previous Section.
More precisely, we study the characteristics of the following set of {\it seventy-six} first order PDEs
\bea
 {\cal T}^{m}&:=&\nabla e^{m}\approx 0\ ,\nn\\[2mm]
 {\cal G}_{m}\, &:=&G_{m}-m^{2}t_{m}\approx 0\, ,\nn\\[2.5mm]
\mathcal{R}^{mn}\!&:=&R^{mn}-d\omega^{mn}-\omega^{m}{}_{t}{{}}\omega^{tn}\approx 0\ .\label{triplet}
\eea
The first forty of these are familiar from the initial set~\eqn{eoms}, while the remaining thirty-six (trivial) equations have been introduced in order to also  treat the Riemann curvature as an {\it independent variable} and thus handle efficiently the curl of the contorsion in the scalar constraint~\eqn{scal}. Hence there are seventy-six dynamical fields $(e^m,\omega^{mn}, R^{mn})$.

We begin our study by assuming the existence of a spacelike, with respect to the dynamical\footnote{One might also choose fiducially spacelike surfaces. This does noes not alter the superluminality conclusions below. Our choice  enables us to also study  dynamical acausalities.} metric~$g_{\mu\nu}$, characteristic surface~$\Sigma$; this can be thought of as the world-sheet of a shock-wavefront propagating at superluminal speeds. More precisely it is characterized as a surface where the first derivatives of the dynamical fields suffer discontinuities in the direction of the normal~$\xi^{\mu}$ to~$\Sigma$; the discontinuity of any quantity~$q$ across this surface will be denoted  by $$\left[\, q\, \right]:= q\, \raisebox{-1mm}{$\big|$}_{{\scriptstyle \Sigma +}}-\ q\, \raisebox{-1mm}{$\big|$}_{{\scriptstyle \Sigma -}}\, .$$
In particular, for the dynamical fields
\beas
{}[\partial_{\mu}e_{\nu}{}^{m}]&:=&\xi_{\mu}\mathfrak{e}_{\nu}{}^{m}\, ,\\[1mm]
{}[\partial_{\mu}\omega_{\nu}{}^{mn}]&:=&\xi_{\mu}\mathfrak{w}_{\nu}{}^{mn}\, ,\\[1mm]
{}[\partial_{\mu}R_{\rho\sigma}{}^{mn}]&:=&\xi_{\mu}\mathfrak{R}_{\rho\sigma}{}^{mn}\ .
\eeas
Since~$\Sigma$ is spacelike, the normal vector obeys $$\xi^{\mu}\xi_{\mu}=-1\, ;$$
there is no loss of generality in normalizing $\xi$.
Throughout  indices will be manipulated using the dynamical metric and vierbein.
The forms~$(\mathfrak{e}{}^{m},\mathfrak{w}{}^{mn},\mathfrak{R}^{mn})$ are the tensors that characterize the shock-wave profile. 

The discontinuities in the field equations~\eqn{triplet}, along with those of their constraints (see the preceding Section), determine 
whether spacelike characteristic surfaces and concomitant superluminalities are permitted:  
Computing the discontinuity across~$\Sigma$ of the equations of motion and gradients of constraints gives a linear homogeneous system of equations in the shock-wave profiles of the form
\be
\raisebox{.5mm}{\scalebox{1.3}{$\chi$}}
\begin{pmatrix}\mathfrak{e}{}^{m} \\\mathfrak{w}{}^{mn} \\\mathfrak{R}{}^{mn}\end{pmatrix}
\approx 0\ .
\label{char1}
\ee
Here~$\chi$ is called the \textit{characteristic matrix}; if it is invertible, space-like characteristics  are excluded.  Note that a field-dependent characteristic matrix usually foretells non-invertibility and thus superluminality.

\subsection{The strategy}

The  characteristic matrix analysis is streamlined by introducing a natural orthonormal basis $(\xi:=\xi_\mu dx^\mu,\varepsilon^{i})$ for the cotangent spaces along the characteristic hypersurface. Any tensorial quantity can be expressed in terms of this basis and its dual, for example a one-form
becomes
$$
\mbox{\resizebox{2.5mm}{2.85mm}{$\theta$}}=-\mbox{\resizebox{2.5mm}{2.85mm}{$\theta$}}_{\!o}\, \xi+\mbox{\resizebox{2.5mm}{2.85mm}{$\theta$}}_i \varepsilon^{i}:= \bm \theta+\mbox{\resizebox{2.5mm}{4.1mm}{$\ring\theta$}}\, .
$$
With these definitions,~$\xi^{o}=1$ and~$\xi^{i}=0$, while~$\xi_{o}=-1$.
Moreover~$g_{oo}=-1$,~$g_{ij}=\delta_{ij}$ and~$g_{oi}=0$. 
\newcommand{\3}{\mbox{\scalebox{.6}{{$\scriptscriptstyle(3)$}}}}
This split into timelike and spatial parts defined by the shock wave-front  allows us to adopt a  notation similar to that of the previous section for differential forms:
any~$p$-form~\resizebox{2.5mm}{2.85mm}{$\theta$} (with~$p<4$) can be decomposed as
\bes
 \mbox{\resizebox{2.5mm}{2.85mm}{$\theta$}}:= \bm \theta+\mbox{\resizebox{2.5mm}{4.1mm}{$\ring\theta$}}\, ,
\ees
where
$$\mbox{\resizebox{2.5mm}{4.1mm}{$\ring\theta$}}\wedge \xi=0\, .$$ Thus $\bm\theta$ is the purely spatial part of the form~$\theta$. 

In the above basis (modulo judicious field redefinitions) the characteristic equation~\eqn{char1} will take the block form
\bes
\left(\begin{array}{c|c|c} \bm 1 & 0 & 0 \\\hline \star & \bm 1& \star \\\hline \star & \star & \star\end{array}\right)\left(\begin{array}{c}\mathfrak{O} \\\hline \mathfrak{T} \\\hline \mathfrak{E}\end{array}\right)\approx 0\ ,
\ees
where $(\mathfrak{O},\mathfrak{T},\mathfrak{E})$ are linear combinations of the shock-wave profiles~$(\mathfrak{e}{}^{m},\mathfrak{w}{}^{mn},\mathfrak{R}^{mn})$.
The  form above  implies that the \textit{null part}~$\mathfrak{O}=0$  and allows us to  solve for the \textit{trivial part}~$\mathfrak{T}$ as functions of the \textit{essential part}~$\mathfrak{E}$. This gives the {\it reduced characteristic equation}
\be\label{rce}
\raisebox{.7mm}{\scalebox{1.1}{$\widehat\chi$}} \, \mathfrak{E}\approx 0\ .
\ee


\subsection{The null and trivial parts}

To obtain the null part of the characteristic equation, our first step is to compute the shock in the equations of motion~\eqn{triplet}:
\beas
[{\cal T}^{m}]\ &=& \xi{{}}\,  \mathfrak{e}^{m}\approx 0\, ,\\[2mm]
[{\cal G}_{m}]\  \, &=&\epsilon_{mnrs}e^n \xi{{}} \, \mathfrak{w}^{rs}\approx 0\, ,\\[2mm]
[{\cal R}^{mn}]&=& \xi{{}}\,  \mathfrak{w}^{mn}\approx 0\, .
\eeas
In general, vanishing of the wedge product of the one-form $\xi$ and a $k$-form $X$ implies $X_{i_1i_2\ldots i_k}=0$; thus
the first and third relation above imply
$$
\mathfrak{e}_i{}^m\approx 0 \approx \mathfrak{w}_i{}^{mn}\, ,
$$
while the second gives no new information. There are further relations contributing to the null part of the characteristic equation; to obtain those, 
we first notice that the curl of the trivial equation in~\eqn{triplet} gives
$
\nabla {\cal R}^{mn} =\nabla R^{mn} \approx 0\, ,
$
whose shock yields $$[\nabla{\cal R}^{mn}]=\, \xi\ \mathfrak{R}^{mn}\approx 0\, ,$$ so that
$$
\mathfrak{R}_{ij}{}^{mn}\approx 0\, .
$$
Further relations on the shock-wave profiles can be obtained by studying the discontinuities in the gradients of the constraints. 
Our analysis is further simplified by using variables that maximize the null part of the characteristic equation. We can indeed
use the variable $f_{\mu\nu}=e_\nu{}^m f_{\mu m}$
 instead of the dynamical  vierbein, so long as the fiducial vierbein is taken to be invertible. 
Calling its shock-wave profile $\mathfrak{f}_{\mu\nu}$, we have
$$
\left[\partial_\mu f_{\nu\rho} \right]:=\xi_\mu\mathfrak{f}_{\nu\rho}\quad\mbox{and}\quad \mathfrak{f}_{\nu\rho}=\mathfrak{e}_\mu{}^m f_{\nu m}\, ,
$$
because all fiducial quantities are assumed to be smooth across $\Sigma$.
Since the symmetry constraint~\eqn{symmetry} then says
$$
{\cal F}_{\mu\nu}=f_{[\mu\nu]}\approx0\, ,
$$
taking the shock of its gradient   we have
$$
-\xi^\mu \left[\partial_\mu {\cal F}_{\mu\nu}\right]=\left[\partial^o{\cal F}_{\mu\nu}\right]=\mathfrak{f}_{[\mu\nu]}\approx 0\, .
$$
In turn, since $\mathfrak{e}_i{}^m\approx 0$, it follows that of the shock-wave profiles $\mathfrak{f}_{\mu\nu}$,  only $\mathfrak{f}_{oo}\, \slashed{\approx}\ 0$.
This has several very useful consequences, in particular $$\mathfrak{f}_{\mu\nu}=\xi_\mu\xi_\nu\mathfrak{f}_{oo}
\quad \mbox{and}\quad
\mathfrak{e}_{\mu\nu}=\xi_\mu l_{o\nu}\mathfrak{f}_{oo}\, ,$$ so that 
$$
\left[\partial^o g_{\mu\nu}\right]=2\mathfrak{e}_{(\mu\nu)}=2l_{o(\mu}\xi_{\nu)}\mathfrak{f}_{oo}\, ,
$$
where $l^\mu{}_m$ is the inverse fiducial vierbein. Hence the shock in the Christoffel symbols is
$$
\left[\Gamma^\rho_{\mu\nu}\right]=\xi_\mu\xi_\nu l_o{}^\rho\mathfrak{f}_{oo}\, .
$$
This allows us to compute the remaining $\mathfrak{R}_{\mu\nu}{}^{mn}$ shock-wave profiles in terms of $\mathfrak{f}_{oo}$. For that we study the shock in the relation~\eqn{divR}. 
Because the shock in the gradient of the  vierbein is proportional to $\mathfrak{f}_{oo}$, the same applies for the shock of the gradient of the mass stress tensor, so we define
$$
\left[\partial^o t_{mn}\right]:=\tau_{mn} \mathfrak{f}_{oo}\, ,
$$
where $t_{m}:=\frac{1}{3!}\epsilon_{nrst}t_{m}{}^{n}e^{r}\wedge e^{s}\wedge e^{t}$.
The tensor $\tau_{mn}$ is easily computed and we find
$$
\tau_{m}{}^n=\frac{1}{2}\epsilon_{mrst}\epsilon^{nlpq}\, {M}_{pq}{}^{st}\xi^{r}\xi_{l}\ ,
$$
where $M_{pq}{}^{mn}=M_{\mu\nu}{}^{mn}e^{\mu}{}_{p}e^{\nu}{}_{q}$ are the components of the two-form $M^{mn}$ in the dynamical vierbein basis. 
Turning to the shock in the relation~\eqn{divR}, we use the above to obtain
\be\label{Rshock}
\mathfrak{R}_{o\nu\rho\sigma}\approx \left[l_{o}{}^{\kappa}\left(R_{\nu\kappa\rho\sigma}+\xi_{\nu}R_{o\kappa\rho\sigma}\right)+2m^{2}\xi_{[\rho}\left(\tau_{\sigma]\nu}-\frac{1}{2}g_{\sigma]\nu}\tau^\kappa{}_\kappa\right)\right]\mathfrak{f}_{oo}\, .
\ee
As a consistency check, one can verify that $\mathfrak{R}_{oo\rho\sigma}=0$  requires
$$
\tau_{o\nu}\approx 0\, ,
$$
which holds because mass stress tensor is weakly conserved, $\nabla^{\mu}t_{\mu\nu}\approx 0$; the shock of this relation gives
precisely the above.
Decomposing the relation~\eqn{Rshock} gives
%
$$
\mathfrak{R}_{oi oj}\approx \left[l_{o}{}^{\kappa}R_{i\kappa oj}-m^{2}\left(\tau_{ij}-\frac{1}{2}g_{ij}\tau^k{}_k\right)\right]\mathfrak{f}_{oo}\quad \mbox{and}\quad
\mathfrak{R}_{oij k}\approx l_{o}{}^{\kappa}R_{i\kappa j k}\mathfrak{f}_{oo}\, .
$$
This completes the determination of the null and trivial parts of the shock-wave profiles. At this juncture, the only independent shock-wave profiles are 
$(\mathfrak{f}_{oo},\mathfrak{w}_{omn})$; these constitute the essential part of the shock-wave profiles and are subject to the reduced characteristic equation~\eqn{rce}.  

\subsection{The reduced characteristic matrix}

To compute the reduced characteristic matrix, we begin by searching for relations on the shock-wave profiles $\mathfrak{w}_{omn}$. These come from
the shocks of the gradients of the curled symmetry  $[\partial^o {\cal K}]\approx0$, and vector  $[\partial^o{\cal V}]\approx0$, constraints. These equations can be written in condensed differential  form notation upon noticing that
$$
\mathfrak{e}^{m}=\ring\mathfrak{e}^{m}=\xi\,  l_{o}{}^{m}\mathfrak{f}_{oo}\quad\text{and}\quad\mathfrak{w}^{mn}=\ring\mathfrak{w}^{mn}=-\xi\, \mathfrak{w}_{o}{}^{mn}\ .
$$
Indeed they are given by
\bea
&l_{o}{}^{m}{\bm K}_{mn}{\bm f}^{n}\mathfrak{f}_{oo}+{\bm e}^{m}{\bm f}^{n}\mathfrak{w}_{omn}\approx 0\ ,
\nn\\[2mm]
&2\epsilon_{mnrs}l_{o}{}^{m}(\beta_{1}{\bm e}^{n}+\beta_{2}{\bm f}^{n}){\bm K}^{rs}\mathfrak{f}_{oo}-\epsilon_{mnrs}{\bm M}^{mn}\mathfrak{w}_{o}{}^{rs}\approx 0\ .\label{greatdane}
\eea
It remains only to shock the scalar constraint~\eqn{scal1}. The key ingredient for this  computation is the Riemann tensor shock~\eqn{Rshock}; we find
\bea\label{scalarshock}
&&\epsilon_{mnrs}\left(\beta_{1}\bm e^{m}\bm e^{t}-2\beta_{2}\bm e^{(m}\bm f^{t)}-3\beta_{3}\bm f^{m}\bm f^{t}\right)\left(\bm K^{nr}\mathfrak{w}_{o}{}^{s}{}_{t}-\bm K^{s}{}_{t}\mathfrak{w}_{o}{}^{nr}\right)\nn\\[2mm]
&&\quad+2\epsilon_{mnrs}\left(\beta_{1}l_{o}{}^{[m}\bm e^{t]}-\beta_{2}l_{o}{}^{(m}\bm f^{t)}\right)\bm K^{nr}\bm K^{s}{}_{t}\mathfrak{f}_{oo}\nn\\[2mm]
&&\quad+2m^{2}\epsilon_{mnrs}l_{o}{}^{m}\Big(4\beta_{0}\beta_{1}\bm e^{n}\bm e^{r}\bm e^{s}+3(\beta_{1}^{2}+2\beta_{0}\beta_{2})\bm e^{n}\bm e^{r}\bm f^{s}\nn\\
&&\qquad\qquad\qquad\quad+6\beta_{1}\beta_{2}\bm e^{n}\bm f^{r}\bm f^{s}+(\beta_{1}\beta_{3}+2\beta_{2}^{2})\bm f^{n}\bm f^{r}\bm f^{s}\Big)\mathfrak{f}_{oo}\nn\\[2mm]
&&\quad-3\epsilon_{mnrs}\beta_{3}\bm f^{m}\bm f^{n}\left(
\bm \rho^{rs}
+2m^{2}\xi^{[r}\Big[\bm \tau^{s]}-\frac{1}{2}\tau^t{}_t \bm e^{s]}\Big]\right)\mathfrak{f}_{oo}\nn\\
&&\quad-4\beta_{2}l_{o}{}^{m}\bar{\bm G}_{m}\mathfrak{f}_{oo} -2\epsilon_{mnrs}\beta_{1}l_{o}{}^{m}\bm e^{n}\bar{\bm R}^{rs}\mathfrak{f}_{oo}\approx 0\ ,
\eea
where the one forms $\rho^{mn}$ and $\tau_m$ are defined by $\rho^{mn}=\rho_{\nu}{}^{mn}dx^\nu:=l_o{}^\mu R_{\mu\nu}{}^{mn}dx^\nu$ 
and 
$\tau_m=\tau_{\mu m}dx^\mu:=$ $\frac{1}{2}\epsilon_{mrst}\epsilon_{\mu}{}^{\nu\alpha\beta}\xi^{r}\xi_{\nu} M_{\alpha\beta}{}^{st}dx^\mu$.

Assembling the system of equations~(\ref{greatdane},\ref{scalarshock}) into 
matrix form determines 
 the $7\times 7$~reduced characteristic matrix $\hat \chi$ as in~\eqn{rce} where the essential part  is $\frak{E}=(\frak{w}^{mn},\frak{f}_{oo})$. This matrix encodes all the necessary information about the well-posedness of the initial value problem for the system of PDEs~\eqn{triplet} and hence also~\eqn{eoms}.

\section{Flat fiducial propagation analysis}\label{pa}

The characteristic matrix is a powerful tool for analyzing the fmGR parameter space
to sort out inconsistent theories. This is because  characteristic surfaces signal a loss of hyperbolicity as well as superluminal shock propagation over a dynamical mean field solution. If our aim were to establish 
complete fmGR consistency, we would have to (i)~calculate the determinant of the  reduced characteristic matrix~$\hat \chi$ determined by equations~\eqn{greatdane} and~\eqn{scalarshock} and (ii) prove that it cannot vanish weakly for any configuration of fields. Of course, one might hope that this singled out a special choice of parameters. This computation is rather involved, and in any case counterexamples
for subsets of the parameter space are already known~\cite{DeffayetBein}. A discussion of how to generally construct zeros of the characteristic determinant, superluminalities and even how to embed closed timelike curves
is given in~\cite{DeserOng}.
Therefore, to illustrate the method, we shall restrict ourselves to 
analyzing some extremely simple physical configurations that already further restrict the allowed parameters.



If we take both background and fiducial metrics flat and aligned, $g_{\mu\nu}=\bar g_{\mu\nu}=\eta_{\mu\nu}$,
a vanishing  characteristic matrix would signal superluminal shocks in the FP theory\footnote{Essentially, the shock in this case is a small perturbation of a continuous Minkowski mean field.}. 
Since the (mean field) contorsions and curvatures vanish, the system~\eqn{greatdane} immediately yields
$$
\frak{w}_{omn}\approx 0\, .
$$
The scalar shock is also extremely simple:
$$\big[m^2(\beta_1+2\beta_2+3\beta_3)\big]^2\, \frak{f}_{oo}\approx0\, .$$
We immediately recognize the left hand side to be $m_{\rm FP}^4$, so non-vanishing FP mass rules out superluminality\footnote{When $m_{\rm FP}=0$, the superluminal modes are pure gauge in the linear FP system.} here.


However, not every field configuration is healthy since {\it a priori} in a theory of a dynamical metric propagating in a fiducial background, the two lightcones
are not guaranteed to be compatible. A simple case is a flat fiducial background and flat dynamical mean field that are not Lorentz-related, for example
$$
d\bar s^2=-dt^2+dx^2 + dy^2+dz^2 \quad \mbox{and} \quad
d s^2=-dz^2+dx^2 + dy^2+dt^2\, .
$$
This configuration solves the equations of motion~\eqn{eoms} iff 
$$
\beta_0+\beta_1+\beta_2+\beta_3=0=\beta_1+2\beta_2+3\beta_3\, .
$$ 
The first condition coincides with the usual one required for the background to solve the equations of motion~\eqn{backsol}, while the second implies vanishing FP mass~\eqn{PauliFierz}.
One might therefore already rule out the parameter choice $\beta_1+2\beta_2+3\beta_3=0$ because
the interacting DoF count does not equal that of the  free (massless spin~2) limit.
We shall instead rule out this theory on grounds of  superluminality.
Consider a putative characteristic constant-$z$ surface\footnote{We label $f^0=dt$, $f^i=dx^i$ and $e^0=dz$, $e^1=dx$, $e^2=dy$ and $e^3=dt$.} (so the normal vector $\xi^\mu\partial_\mu=\frac{\partial}{\partial z}$). 
Then the curled symmetry shock (the first of the equations~\eqn{greatdane}) here implies $\frak{w}_{o12}=0$,
$\frak{w}_{o23}=\frak{w}_{o02}$ and $\frak{w}_{o13}=\frak{w}_{o01}$. In turn, the vector constraint shock implies
$$
(\beta_1+2\beta_2+3\beta_3)\frak{w}_{o0i}\approx 0\, .
$$ 
Since the equations of motion already imply the vanishing coefficient of the shock wave-fronts~$\frak{w}_{o0i}$ in the above, these are not determined so the characteristic matrix has a vanishing determinant, and the model is indeed superluminal. 

We can  also probe whether  spacetimes in which the fiducial and dynamical metrics have different speeds of light can lead to superluminalities.  For this we take a dynamical metric ansatz, $$ds^2=-c^{2}dt^2+dx^2+dy^2+dz^2\, ,$$ and a   Minkowski fiducial metric. Again this configuration solves the equations of motion \eqn{eoms} iff
$$\beta_{0}+\beta_{1}+\beta_{2}+\beta_{3}=0=(\beta_{1}+2\beta_{2}+3\beta_{3})(c-1)\ .$$
Since $c=1$ reduces to the previous FP situation, we consider $c\neq1$ (so that  $m_{\rm FP}$ must vanish) and then study a constant-$t$ putative characteristic surface with $\xi^\mu\partial_\mu=\frac{\partial}{\partial t}$.
Since both metrics are flat, we note that all contorsions vanish, and $f_o{}^i=0$. Then the shocked curl of the symmetry constraint becomes a homogeneous, trivially invertible system, forcing~$\mathfrak{w}_{oij}=0$.  The vector constraint's shock now reduces to 
$$(\beta_1+2\beta_2c+3\beta_3c^{2})\, \mathfrak{w}_{ooi}\approx0\, .$$ 
Clearly, for generic $\beta$s, there are values of $c$ such that the coefficient above is zero, hence  we already detect superluminalities with non-zero shock-wave profile $\mathfrak{w}_{ooi}$. Alternatively,  keeping~$c$ generic, there then exists some combination of the $\beta$s such that  the coefficient of~$\mathfrak{f}_{oo}$ in the shock of the scalar constraint~\eqn{scalarshock} vanishes. In other words, we can find models with superluminality for any value of $c$.

Analysis of more complicated solutions with non-flat fiducial backgrounds will harness the full power of the reduced characteristic matrix calculated in the previous section, but these simple examples already demonstrate the mechanism responsible for superluminal propagation.  Introducing more general  fiducial  field dependence will generically only make  matters worse. 


%

\section{Conclusion}\label{c}

We have performed a definitive analysis of full generic fmGR's propagation properties. By employing a first order Palatini formalism, 
we were able at last to obtain the explicit covariant constraints' form,   valid for the theory's full parameter space. This result then enabled us to  compute the characteristic matrix for all parameters. A new feature here, in the  hitherto unprobed third parameter direction, is
the appearance of the full dynamical Riemann curvature in the field-dependent characteristic matrix (which,  in previous studies that were limited to subsets of parameter space, only involved metrics/vierbeine and contorsions). 

The characteristic matrix is a powerful tool for analyzing any theory. In particular, if it is field-dependent there are many potential difficulties. It is intimately related to the kinetic structure and hence canonical commutators of the quantum version of the theory. Hence, a degenerate characteristic matrix implies zero and negative norm states~\cite{KS}. It also determines characteristic surfaces, where  predictability is lost  and along which shocks propagate. Thus spacelike characteristic surfaces are very dangerous for any model. It is even possible to use them to detect micro-acausality  (local CTCs). All these pathologies\footnote{While we have discussed matter couplings, it should be noted that they present additional problems~\cite{DWmass}. Nor have we considered fmGR's  strong coupling pitfalls~\cite{Nemanja}.} are known features of fmGR~\cite{DeserOng,DW,DSW,Gr}. 

Our main causality result is the characteristic matrix itself, which encodes all this information.   There is one last fmGR glimmer of hope, namely that for some distinguished choice of fiducial background and parameters, the characteristic matrix could be non-degenerate. This seems highly unlikely, since already counterexamples are known for broad classes of backgrounds and field configurations (see~\cite{DeserOng}), the very simplest examples of which were exhibited in section~\ref{pa}.
These showed not only how easy it is to construct pathological solutions but also why models depending on fiducial backgrounds
lead to an enormous loss of predictability: Even supposing that the model has a limited viability as an effective theory, one would have to first choose a background {\it by hand} and then check that it supports well-defined propagation for the spacetime region being considered.
Without a principle for choosing the background, observational predictability is clearly imperiled.

Another issue, to which we gave little focus, is that for some regions in parameter space, the model has different branches~\cite{DeffayetBein}
because the symmetry constraint is not the unique solution to~\eqn{symmetry} (in a second order metric formulation there is a similar issue related to the existence of square roots of the endomorphism $g_{\mu\nu}\bar g^{\nu\rho}$ used to define the mass term). Models with branches can suffer both 
jumps in DoF counts and loss of unique evolution. 

The above  list of fmGR pathologies suggests that a possible panacea could be the original bimetric model where the fiducial metric is dynamical~\cite{Salam}. A characteristic analysis for the bimetric theory is currently unavailable, but since the causal structures of two dynamical metrics are guaranteed to conflict with one another, there seems little hope for consistency here either. Also, various studies
have indicated that the bimetric theory possesses no partially massless limit~\cite{PMbi} (even though its linearization does~\cite{HassanPM}), which is strong evidence against models of this type. 

\section*{Acknowledgements}
The work of S.D. was supported in part by NSF PHY-1266107 and DOE  DE-SC0011632 grants and that of G.Z. in part by
DOE Grant DE-FG03-91ER40674. 
We thank C. Deffayet,   K. Hinterbichler, K. Izumi, N. Kaloper, J. Mourad and Y.C. Ong for discussions.

\appendix

\section{The Fierz--Pauli limit}\label{A}

To check that the  thirty constraints found in Section~\ref{cca} are independent,
we review why this at least holds true in their linear limit.  We first expand the  dynamical fields  around fiducial ones, $$e^m:=f^m+h^m\, ,\quad\omega^{mn}:=\overline{\omega}^{mn}+K^{mn}\, ,$$ where the background is Einstein:
$$
\bar G_m = \frac1{3!}\, \bar \Lambda\,  \epsilon_{mnrs} f^n f^r f^s\, .
$$
Here $\bar G_m:=\frac12\epsilon_{mnrs}f^n \bar R^{rs}$ 
and  $\bar \omega^{mn}$ encodes the fiducial Levi-Civita connection $\bar \nabla$. As already discussed, for this background to be a solution, we must require that
$$
\frac{\bar \Lambda}{3!} = m^2 \left(\beta_0+\beta_1+\beta_2+\beta_3\right)\, .
$$
The linearized equations of motion are then
\bea\label{leoms}
\frac12\epsilon_{mnrs}f^n\bar\nabla K^{rs}&\approx&m^2\, \left[3\beta_0+2\beta_1+\beta_2\right]\, \epsilon_{mnrs}\,  f^nf^rh^s\, ,\nn\\[3mm]
\bar\nabla h^m+f_nK^{nm}&\approx&0\, .
\eea

Just as for their  non-linear counterparts~\eqn{eoms}, the spatial parts of the above field equations yield sixteen primary constraints.
Next we find six secondary constraints from symmetry of the linearized Einstein tensor, and a further four secondary constraints by computing the curl of the first equation in~\eqn{leoms}
(the linearized vector constraint)
\bes
f^m h_m\approx0\approx \frac34\, m_{\rm FP}^2\, \epsilon_{mnrs}f^mf^n K^{rs}\, .
\ees
Here we have defined, as earlier, the FP mass, 
$$m_{\rm FP}^2:=m^2(\beta_1+2\beta_2+3\beta_3)\, .$$ 
The  linearized vector constraint becomes an identity precisely at $m_{\rm FP}=0$, where the model describes massless gravitons.

 The remaining four constraints are tertiary and found from the curls of the secondary constraints:
$$
f^m f^n K_{mn}\approx 0 \approx \frac12 \, m_{\rm FP}^2(\bar\Lambda-\frac32 m_{\rm FP}^2) \epsilon_{mnrs} f^m f^n f^r h^s\, .
$$
Notice that at the value $m_{FP}^2=\frac23\bar\Lambda$, the last---linearized scalar---constraint is elevated to a gauge invariance.  Indeed this is precisely the partially massless tuning found in~\cite{PM}, an invariance is known not to survive in the full nonlinear theory~\cite{DSW,deRhamPM}. Finally, as promised, observe that all constraints are independent.
In particular, they imply that, for $\{m^2_{\rm FP}\neq 0,\frac23\bar\Lambda\}$, the dynamics are described by a symmetric tensor $h_{\mu\nu}$ that is (fiducially) trace- and divergence-free.

%

\section{Hamiltonian Analysis}\label{B}

We now give an  account of the Hamiltonian analysis of fmGR in Palatini formalism\footnote{See~\cite{Banados} for a three-dimensional Palatini-based fmGR Hamiltonian analysis.}. This computation was first performed for pure gravity in~\cite{DeserIsham} (see also~\cite{Charap}).
Writing the fmGR action as an integral~$S=\mbox{\scalebox{1.1}{$\int$}} {\mathscr L}$ over a sum of volume forms~${\mathscr L}:= {\mathscr L}_{\rm GR}+{\mathscr L}_{\rm fm}$  where
\begin{equation}\label{eew}\begin{split}
{\mathscr L}_{\rm GR}&:=-\frac{1}{4}\,  \epsilon_{mnrs}\,  e^m{{}} e^n \, \left[d\omega^{rs} + \omega^r{}_t{{}}\omega^{ts}\right]\, ,\\[2mm]
{\mathscr L}_{\rm fm}\, &:=m^2  \epsilon_{mnrs}\, e{}^m{{}}\left[\frac{\beta_{0}}{4}e^n{{}} e^r{{}} e^s +\frac{\beta_{1}}{3}e^n{{}} e^r{{}} f^s+\frac{\beta_{2}}{2}e^n{{}} f^r{{}} f^s+\beta_{3}f^n{{}} f^r{{}} f^s\right]\, ,
\end{split}
\end{equation}
our first task is to decompose these into a $3+1$ split. For that, we employ the notations of Section~\ref{cca} and define
$$\mathscr L=:dt \wedge \bm L \, ,$$
with
~${\bm L}:={\bm L}_{\rm GR}+{\bm L}_{\rm fm}$.
In addition we call
$$
\mathring{e}^m=:dt \, N^m\, ,\quad \mathring{f}^m=:dt \, \bar N^m \quad\mbox{and}\quad \mathring\omega^{mn}=:dt\,  w^{mn}\, .
$$
Then (up to surface terms)
$$
{\bm L}=-\frac14 \epsilon_{mnrs}\,  {\bm e}^m {\bm e}^n \, \dot {\bm \omega}^{rs} - N^m \bm{\mathcal G}_m
-\frac12w^{mn} \epsilon_{mnrs}\bm{\mathcal T}^{\, r}\bm e^s-{\bm H}(\bm e^m)\, ,
$$
where $\bm{\mathcal G}_m$ and $\bm{\mathcal T}^m$ are defined in equation~\eqn{frogs} and
$$
{\bm H}(\bm e^m)=
m^2  \epsilon_{mnrs}\, \bm e{}^m{{}}\left[\frac{\beta_{1}}{3}\bm e^n{{}} \bm e^r{{}} +\beta_2\bm e^n{{}} \bm f^r{{}} +3\beta_{3 }\bm f^n{{}} \bm f^r{{}} \right]\bar N^s\, .
$$
%
Upon integrating the time derivative by parts, the action has twelve canonical pairs,~${\bm e}^m$ and their conjugate momenta~$\frac12\epsilon_{mnrs}\,  {\bm e}^n \, {\bm \omega}^{rs}$. These are ostensibly subject to ten constraints imposed by the Lagrange multipliers~$N^m$ and~$w^{mn}$. For  GR, it was   shown that 
this model is equivalent, upon integrating out the six $\omega^{mn}$,  to the  standard ADM form involving six canonical pairs built from spatial metrics and their momenta, but still subject to four diffeomorphism constraints~\cite{DeserIsham}, thus yielding two physical DoF. 
For our purposes, however, instead of returning to a metric-based ADM formulation, it is advantageous to decompose the model such that its dependence on the spatial dreibeine~${\bm e}^a$ (splitting flat indices  
$
m=(0,a)
$) manifests three-dimensional coordinate and Lorentz invariance.
To that end, recall that the Einstein--Hilbert action~\eqref{eew}
takes its familiar~$S_{EH}=\mbox{\scalebox{1.1}{$\int$}} \sqrt{-g}\,  R$ form upon algebraically integrating out the spin connection {\it a la} Palatini, by solving the torsion constraint
\begin{equation}\label{torsion}
0=de^m +\omega^{m}{}_n e^n\, .
\end{equation}
Instead of integrating out the entire~$\omega^{mn}$, which would also return us to the metric ADM formulation,  we solve the above condition only for the spatial spin connections\footnote{This parallels the canonical analysis of ``Palatini'' Maxwell theory: there $\bm B$ is solved for in terms of $\bm d \bm A$, but $\bm E$ is kept independent. }. This is achieved by further decomposing  the dynamical fields according to 
$$
{\bm e}^m:=({\bm M},{\bm e}^a)\, , \quad \ring e^m:= (N,N^a)\, dt\, ,\quad
{\bm \omega}^{0a}:={\bm P}^a\, , \quad \ring\omega^{0a}:=u^a\, dt\, .
$$
The variables $N$ and $N^a$ correspond to  ADM lapse and shift variables. We will call $\bm M:=M_i dx^i$ the {\it shaft} while $\bm P^a:=P_i{}^a dx^i$ will become nine canonical momenta. 
Thus we must now solve nine of the twenty-four torsion constraints:
$$
0={\bm d} {\bm e}^a + {\bm \omega}^a{}_b {\bm e}^b + {\bm P}^a {\bm M}\, .
$$
To that end, we henceforth assume invertibility of the dreibein $\bm e^a:=e_i{}^a dx^i$ and use it to manipulate three-dimensional indices. The torsion solutions are then 
$$
{\bm \omega}^{ab}({\bm e,\bm M,\bm P})={\bm \omega}^{ab}(\bm e) + M^{[a} {\bm P}^{b]}-{P}^{[ab]} {\bm M}-{P}^{[a}{}_c {M}^{b]} {\bm e}^c=:{\bm \omega}^{ab}(\bm e)+{\bm K}^{ab}\, .
$$
 From now on, we   denote the three-dimensional Levi-Civita connection based on the Levi-Civita spin connection $\bm \omega(\bm e)$ by  $\bm\nabla$,  and $\bm \nabla_{\!\sss\bm K}$ is
 its torsionful counterpart based on the spin connection in the above display (do not confuse $\bm K$ with the analogous quantity introduced earlier). 
 The respective curvature two-forms will be denoted by ${\bm R}^{ab}$ and ${\bm R}_{\sss\bm K}^{ab}$. Substituting this solution into the lagrangian ${\bm L}$
 yields
\begin{equation}\label{3+1}
{\bm L}={\bm P}_a 
\dot{\tilde{\bm e}}^a
+\frac12\epsilon_{abc}{\bm M}{\bm e}^a \dot {\bm \omega}^{bc}({\bm e,\bm M,\bm P})
-N \bm{\mathcal G}_0 - N^a \bm{\mathcal G}_a -\frac12  w^{ab} \bm {\mathcal J}_{ab} -{\bm H}(\bm e,\bm M)\, ,
\end{equation}
where 
\begin{align*}
\bm {\mathcal G}_0 \ &=-\frac12\epsilon_{abc}\big({\bm e}^a [{\bm R}^{bc}_{\sss\bm K}+\bm P^b \bm P^c]-2m^2[\beta_0\bm e^a \bm e^b \bm e^c+\beta_1 \bm e^a \bm e^b \bm f^c 
+\beta_2 \bm e^a \bm f^b \bm f^c +\beta_3 \bm f^a \bm f^b \bm f^c 
]\big)\, ,\\[1mm]
\bm{\mathcal  G}_a\ &=-\frac12\epsilon_{abc}\Big(2{\bm e}^b\bm \nabla_{\!\sss\bm K} \bm P^c -\bm M\big[{\bm R}^{bc}_{\sss\bm K} +\bm P^b\bm P^c-2m^2(3\beta_0\bm e^b\bm e^c+2\beta_1 \bm e^b \bm f^c+\beta_2\bm f^b \bm f^c)\big]\\&\qquad\qquad\qquad\quad\ \, 
+2m^2\bar {\bm M}\big[
 \beta_1 \bm e^b \bm e^c+2\beta_2\bm e^b \bm f^c
+3\beta_3 \bm f^b \bm f^c\big]
\Big)\, ,\\[1mm]
\bm {\mathcal J}_{ab} &=\tilde{\bm e}_{a} {\bm P}_{b}-\tilde{\bm e}_{b} {\bm P}_{a}  - \epsilon_{abc}\big[{\bm e}^c\,  {\bm d} {\bm M} -(\bm \nabla_{\!\sss\bm K} \bm e^c)\,  \bm M\big]\, ,\\[1mm]
{\bm H}\ \,  &=-m^2  \epsilon_{abc} \left[\bm M\bm M^{ab} \bar N^c\!- \bm e^a \Big(
\frac{\beta_{1}}{3}\bm e^b{{}} \bm e^c{{}} +\beta_2\bm e^b{{}} \bm f^c{{}} +3\beta_{3 }\bm f^b{{}} \bm f^c{{}} 
\Big)\bar N
+\bm e^a\left(\beta_2\bm e^b + 6\bm f^b \right)\bar{\bm M} \bar N^c\right] 
.
\end{align*}
Here we have introduced the two-form~$\tilde{\bm e}_a:=-\frac12 \epsilon_{abc} {\bm e}^b {\bm e}^c$ (where~$\epsilon_{abc}:=\epsilon^0{}_{abc}$), which may equivalently be viewed as the dual of the (densitized) inverse dreibein; this relation may be inverted for~$\bm e^a(\tilde {\bm e})$. For the fiducial vierbein, we have defined $\ring f^m:=\bar N^m=(\bar N,\bar N^a)$ while $\bar{\bm M}$ is the fiducial shaft. Also
note that the triplet of auxiliary fields~$u^a$ completely decouples  because nine of the torsion constraints have been solved.

Equation~\eqref{3+1} is the key to our Hamiltonian analysis: The quartet of auxiliary fields~$N^m=(N,N^a)$ 
 play the {\it r\^ole} of the shift and lapse Lagrange multipliers in  standard ADM, and we shall henceforth so refer to them. The first term in~\eqref{3+1} is the Darboux form for nine canonical pairs~$(\tilde{\bm e}_a,\bm P^a)$; however, this is complicated by the presence of the second term that potentially involves 
time derivatives of the dreibeine, shaft and canonical momenta. In Einstein gravity, this difficulty is easily circumvented by using a local diffeomorphism  to gauge away the shaft~\cite{DeserIsham}. In an fmGR setting, that route is closed to us; instead therefore, we integrate out the shift Lagrange multipliers $N^a$. This imposes three relations
$\bm {\mathcal G}^a =0$, which we can {\it generically} solve for the three components of the shaft ${\bm M}={\bm M}(\bm e,\bm P)$.
Configurations where these relations do not determine the shaft are, of course,  intimately related to the model's propagation difficulties.
Hence, at this point~$\bm \omega^{ab}=\bm \omega^{ab}(\bm e,\bm P)$ and the Lagrangian becomes
$$
{\bm L}={\bm P}^a\dot {\tilde{\bm e}}_{a}
+\frac12{\bm M}(\bm e,\bm P)\epsilon_{abc}{\bm e}^a \dot {\bm \omega}^{bc}({\bm e,\bm P})
-N \bm {\mathcal G}_0(\bm e,\bm P)  - \frac12 w^{ab} \bm {\mathcal J}_{ab}({\bm e,\bm P}) -{\bm H}(\bm e,\bm P)\, .
$$  
Given that the symplectic terms now depend only on $(\bm e^a,\bm P^a)$
the model has maximally nine canonical pairs.  The three Lagrange multipliers~$w^{ab}$ impose three constraints, which ought remove three of 
these pairs (these constraints are algebraically solvable\footnote{Strictly, this is manifestly the case before setting~$\bm M=\bm M(\bm e,\bm P)$, but we could have first integrated out~$w^{ab}$ and then the shift, so the only real solvability concern is in that constraint.} for the antisymmetric part of the canonical momenta~${\bm P}_{[ij]}$).
The model will then  reduce to six canonical pairs, subject to the single ``Hamiltonian''\footnote{Here we use the modifier Hamiltonian to refer to the timelike diffeomorphism constraint of GR in an ADM approach; the massive model, of course,  still has a non vanishing Hamilitonian~${\bm H}(\bm e,\bm P)$ on the constraint surface.} constraint $\bm {\mathcal G}_0=0$ imposed by the lapse Lagrange multiplier. This computation (modulo checking that the secondary constraint structure is correct) thus shows in 3+1 form that the model generically describes five DoF.

\end{document}